\documentclass[reprint,pre,amsmath,amssymb,aps,floatfix,superscriptaddress]{revtex4-2}
\usepackage{graphicx,bm,amsmath,amsfonts}
\usepackage{comment}
\usepackage{hyperref}
\usepackage{color,xcolor,comment}
\usepackage{amssymb} 
\usepackage{tikz}
\newcommand{\tg}[1]{{\textcolor{black}{#1}}}

\begin{document}

\title{In vivo measurements of fascia lata {effective} mechanics combined to a memory fiber–recruitment–viscoelastic modeling approach}

\author{Franck Germain}
\affiliation{Kinéquipe, Maison de santé, 24 rue Carnot, 39200 St Claude, France}
\author{Thomas Gibaud}
\affiliation{ENSL, CNRS, Laboratoire de physique, F-69342 Lyon, France}
\affiliation{ Department of Polymer Engineering, IPC,University of Minho, Guimarães, 4804-533 Portugal}

\begin{abstract}
{The fascia lata plays a central role in force transmission and body mechanics, yet its in vivo mechanical behavior remains poorly characterized. Existing approaches --- shear wave elastography and direct force measurements alike --- share a fundamental limitation: none simultaneously captures both the elastic and viscous components of fascial mechanics within a single experiment. The primary aim of this study is therefore to develop an experimental and modeling framework that enables the reproducible measurement of the effective viscoelastic properties of the fascia lata in vivo. To this end, we combine controlled ramp--relaxation experiments on the human fascia lata with a constitutive model that integrates fiber recruitment and dual-timescale viscoelastic relaxation. We emphasize that this is an effective model: rather than describing intrinsic local material properties, it characterizes the mechanical response of the fascia lata complex including its coupling to the hip--thigh musculoskeletal system under controlled loading conditions. The model 
captures both the nonlinear stiffening during elongation and the dual decay of force during relaxation, using a minimal set of physically interpretable parameters. Repeated trials demonstrate good reproducibility, with parameter variability within $10\%$. Our results support the view that fascia lata behaves as a hierarchical, hydrated composite whose macroscopic mechanical response emerges from the coupled effects of collagen alignment, matrix viscoelasticity, and fluid flow. This work provides a quantitative foundation for future in vivo investigations into how training, rehabilitation, or aging influence the evolution of fascial mechanical properties.}
\end{abstract}

\maketitle

\section{Introduction}

The fascia lata is one of the longest collagenous structures in the human body~\cite{otsuka2018,otsuka2020} and plays a central role in musculoskeletal function~\cite{eng2015a}. Acting as a tensioned connective sheath, it transmits forces between muscles and bones, stabilizes the hip and knee through the iliotibial tract, and contributes to locomotor efficiency through elastic energy storage and dissipation. Its collagenous and viscoelastic structure enables smooth gliding between muscle layers, preserves flexibility and range of motion, and supports hydration and interstitial fluid flow via its glycosaminoglycan-rich matrix, which facilitates mechanical recovery and maintains viscoelastic properties during movement.

The fascia lata is implicated in a broad spectrum of musculoskeletal conditions, including iliotibial band syndrome and patellofemoral pain syndrome~\cite{merican2009iliotibial}, and plays a key role in post-surgical recovery following tendon or ligament autograft procedures~\cite{eng2014,bonaldi2023}. Beyond pathology, the fascia lata is increasingly recognized as a mechanically adaptive structure whose viscoelastic properties may evolve in response to training~\cite{germain2016mecanismes,germain2020,warneke2024effects}. Quantifying such adaptations in living subjects is therefore of direct clinical relevance: it would enable clinicians and researchers to objectively monitor fascial remodeling throughout rehabilitation or athletic conditioning, and to optimize intervention timing and intensity on an individual basis~\cite{bonaldi2024,stecco2009}. Yet the overwhelming majority of quantitative mechanical characterizations of the fascia lata have been performed \textit{ex vivo}, on cadaveric or fresh-frozen samples~\cite{otsuka2018,bonaldi2023,pancheri2014,eng2014,szotek2024,bonaldi2024}. As recently highlighted in a systematic gap analysis of the field~\cite{bonaldi2024}, the biomechanical literature on the fascia lata is almost entirely restricted to such experimental configurations, which cannot replicate the physiological hydration state, active perfusion, in situ boundary conditions, or the multiaxial muscle-driven loading that characterize the living tissue.

In vivo mechanical characterization of the fascia lata remains scarce. Germain and colleagues~\cite{germain2016mecanismes,germain2020} designed a direct 
force--elongation experiment and demonstrated that static and sustained stretching leads to significant increases in stretch tolerance and viscosity-driven adaptation, indicating that the fascia lata behaves as a viscoelastic structure capable of gradual accommodation to mechanical load. These studies, however, lack quantitative modeling and cannot disentangle the elastic and viscous contributions to the measured response. Beyond direct mechanical testing, the main non-invasive in vivo tool available is shear wave elastography~(SWE), which estimates tissue stiffness from the propagation speed of acoustic shear waves. SWE provides a scalar shear modulus at a single, discrete strain state, typically under active muscle contraction~\cite{otsuka2020}, and yields no information on nonlinear mechanical behavior or viscoelastic relaxation timescales. Warneke~\emph{et al.}~\cite{warneke2024effects} represent a first attempt to detect stretch-induced stiffness changes in deep fascia using strain elastography; however, his approach is semi-quantitative, restricted to acute elastic measurements, and does not yield constitutive model parameters suitable for longitudinal tracking. Furthermore, and somewhat ironically, SWE is itself subject to a tissue-mixing problem at least as severe as that of direct mechanical testing: because the shear wave propagates through all overlying layers --- skin, subcutaneous fat, and muscle --- the measured modulus reflects a composite response rather than an intrinsic fascial property. Taken together, existing in vivo approaches share a common and critical limitation: none simultaneously characterizes both the elastic and viscous components of fascial mechanics within a single experiment. Yet these two contributions are physically coupled --- the viscoelastic relaxation depends on the loading history accumulated during the elastic phase --- so that separating them across independent experiments inevitably introduces inconsistencies and precludes a full constitutive description.

From a physical perspective, the mechanical behavior of the fascia lata emerges from its hierarchical organization across scales~\cite{szotek2024,bonaldi2024}. At the fiber scale, collagen bundles are arranged in a characteristic crimped configuration that progressively straightens under tension, producing the nonlinear, strain-stiffening response observed experimentally~\cite{szotek2024,bonaldi2024}. At the molecular scale, these fibers are embedded in a glycosaminoglycan~(GAG)-rich matrix whose hydrated, gel-like nature governs viscous dissipation through two broadly separable mechanisms: rapid local water redistribution and fibril--matrix sliding~\cite{screen2011,gupta2009}, and slower large-scale fluid exchange and matrix reorganization~\cite{gupta2009,munster2013,nam2016}. This multiscale structure naturally leads to a viscoelastic response involving multiple timescales.

{Existing constitutive models capture parts of this behavior but remain incomplete. Fascia-specific models have been developed primarily from \textit{ex vivo} experiments, with an emphasis on nonlinear elasticity and anisotropy. Stecco \emph{et al.}~\cite{stecco2009} proposed a layered, fiber-reinforced, crimped model that successfully captures the nonlinear elastic response under tension. Bonaldi \emph{et al.}~\cite{bonaldi2023} reproduced directional stiffness and stress--strain curves with a model accounting for intra- and inter-layer interactions. Pancheri \emph{et al.}~\cite{pancheri2014} developed a bi-layered hyperelastic model validated under uniaxial and planar biaxial extension. While these models successfully capture elastic anisotropy, they share important limitations: viscoelastic phenomena such as relaxation, creep, and hysteresis are either omitted or simplified; sample preparation alters hydration and matrix behavior; and in vivo loading conditions --- multiaxial, dynamic, with fluid flow and muscle coupling --- are not represented. More generally, constitutive models for collagenous soft tissues fall into two broad categories. Microstructural and homogenization approaches~\cite{lanir1980microstructure,ganghoffer2016nonlinear,freed2016viscoelastic} account for fibril orientation and matrix interactions, but involve a large number of parameters that are difficult to identify from in vivo data. Recruitment-based models~\cite{romero1998recruitment,raz2009recruitment,bevan2018biomechanical} offer a more tractable alternative by emphasizing the progressive engagement of collagen fibrils under tension, successfully reproducing the toe region of the stress--strain curve with fewer, physically interpretable parameters. Nevertheless, they generally treat fibril--matrix interactions and GAG-mediated fluid effects in a simplified manner, and typically treat viscous dissipation as independent of the recruited fiber fraction. Across both categories of models, a common and fundamental limitation therefore emerges: the coupling between fiber recruitment and viscoelastic dissipation is either absent or treated additively. This decoupling is physically inconsistent --- only taut, load-bearing fibers generate fiber--matrix drag and participate in osmotic fluid exchange --- and has a direct consequence for ramp--relaxation experiments: because viscoelastic elements accumulate internal stress \textit{during} the recruitment process, the relaxation response encodes the full rate and extent of the preceding stretch. No existing model simultaneously accounts for this strain-dependent viscosity, the resulting mechanical memory across the stretching--relaxation transition, and the dual-timescale dissipation arising from the GAG-rich matrix --- three features that are physically inseparable and clinically unavoidable in any in vivo characterization of the fascia lata. What is therefore needed is a minimalist model that couples fiber recruitment multiplicatively to at least two viscoelastic elements operating on well-separated timescales, is parsimonious enough to be fitted to sparse in vivo data, and yields physically interpretable parameters suitable for longitudinal monitoring.}

{The primary aim of this study is to develop such an experimental and modeling framework for the reproducible measurement of the effective viscoelastic properties of the fascia lata \textit{in vivo}. Owing to the mechanical coupling between the fascia and surrounding tissues inherent to in vivo conditions, the measured force reflects the global response of the hip--thigh system rather than a purely intrinsic material property. We therefore seek to define a robust, protocol-dependent mechanical signature under controlled loading conditions. In vivo full characterization of fascial viscoelastic properties remains challenging because truly intrinsic fascial properties are difficult, if not impossible, to isolate in the living subject. Nevertheless, echography has shown that the fascia lata is selectively stretched by the combination of hip extension and knee flexion~\cite{umehara2015effect}, and it has been established that the viscoelasticity of the rectus femoris is not significantly altered by stretching~\cite{nakamura2021comparison,kranjc2025acute}. This low muscular contribution is further supported by a mechanical analysis of anterior thigh stretching~\cite{germain2024mechanical} and by the demonstration that anterior thigh stretch tolerance depends more on the taut surfaces of the fascia lata than on the rectus femoris moment arm at hip level~\cite{germain2023stretch}. We therefore designed a force--elongation setup based on~\cite{germain2020,germain2016mecanismes} that minimizes contributions from surrounding structures as much as anatomy permits; the quantities we access are effective mechanical properties dominated by the fascial response, and the associated measurement uncertainty is explicitly quantified. Although this description is not local, it is reproducible and physically interpretable, and thus provides a quantitative baseline to monitor the evolution of fascia lata effective mechanics over time --- for instance during rehabilitation or in response to training.}

{A key originality of our approach lies in the experimental protocol itself: a ramp elongation immediately followed by a relaxation phase in a single continuous test captures both the elastic response and the viscoelastic relaxation --- including mechanical memory accumulated during the loading ramp --- simultaneously and in the same subject. To extract physically interpretable parameters from these experiments, we introduce a minimalist constitutive model that couples a fiber recruitment function to two viscoelastic Maxwell elements operating on distinct timescales, enabling the simultaneous description of nonlinear elasticity, strain-dependent viscosity, and history-dependent relaxation using a small number of parameters. We demonstrate that the model accurately reproduces both the stretching and relaxation phases, and that the extracted parameters are reproducible within approximately $10\%$ across repeated trials and varying stretching protocol such as the velocity of the stretching $v$, and the maximum force reached during stretching $F_0$. Finally, we discuss the physical values of the viscoelastic fitted parameters in relation to the literature and propose possible associations between the macroscopic model parameters and the underlying nanoscale mechanisms.}

\section{Materials and Methods}
\subsection{Participants and Ethical Approval}

Before the clinical investigation, a 47 year-old physiotherapist assessed the apparatus used to stabilize the pelvis during the stretch. This step was necessary to check pelvic stability during a long stretching session. These preliminary assessments were conducted to validate the experimental setup prior to a full clinical study. The protocol was submitted to the ethical review board of Nice Medical University and was officially registered as N° 19.11.03.77151.
 
The participant was 182 cm and 82 kg. He used to practice cross country skiing, sport climbing, and cycling. He was used to stretch. His right leg was the dominant one and his left medial portion of his fascia lata was injured in a rock fall when he was 16.   

\begin{figure*}
    \centering
    \includegraphics[width=0.99\textwidth]{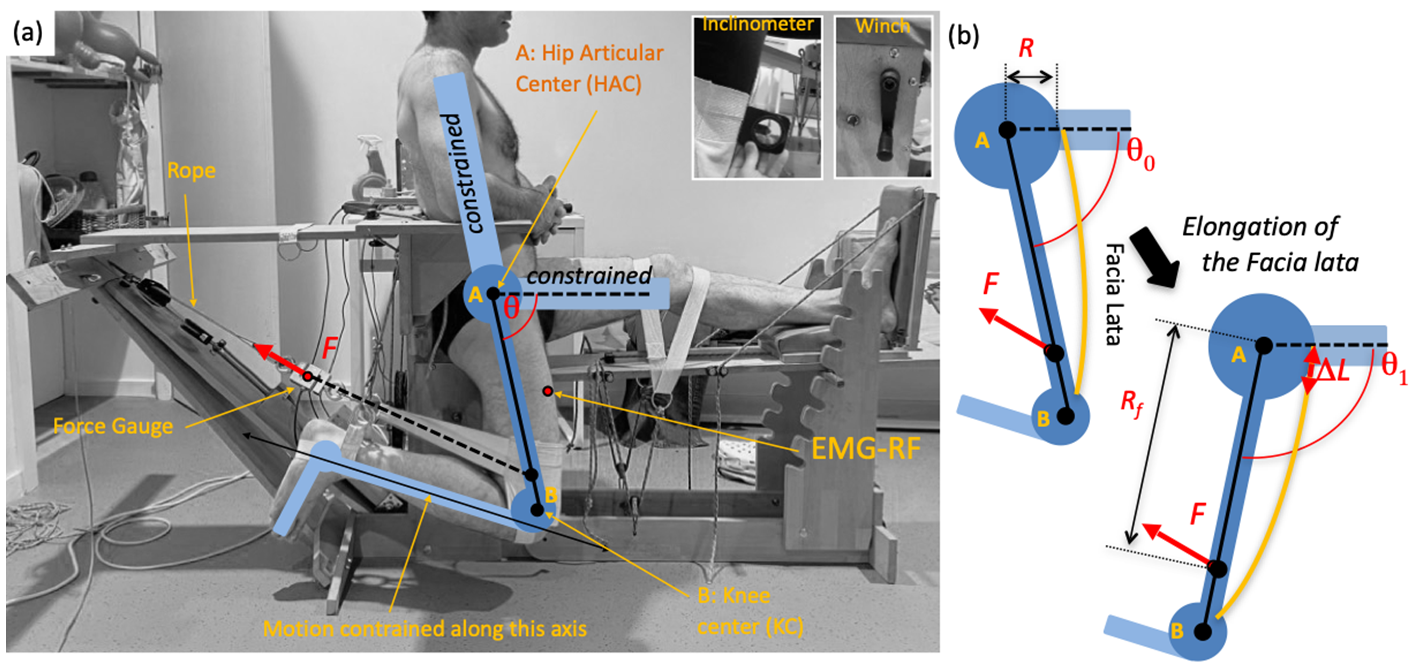} 
    \caption{Experimental Setup. (a) Image of the setup. The pelvis is immobilized to ensure the hip articular center (HAC) remains stationary during measurements.  The degree of hip extension is adjusted using a rope controlled by an actuator. The reaction force of the fascia lata $F$ is recorded over time using a force gauge. An inclinometer and a winch were used to check that the angle ($\theta_i$) was proportional to the travel of the rope determined by the winch or the actuator. The surface electromyogram of the rectus femoris (EMG RF) is recorded using a biofeedback apparatus, with voluntary contractions detected via a cursor set at 2~$\mu$V. During force measurements, data are selected where the EMG RF signal corresponds to a resting activity level. (b) Schematic of the elongation of the fascia lata. The radius of the fascia lata from the hip articulation center $R$ is approximately 6~cm. The radius between the thigh sling and the hip articulation center $R_f$ is approximately 30~cm. The elongation of the fascia lata $L$ is measured using the travel of the rope induced by the actuator.}
    \label{fig1:exp}
\end{figure*}

\subsection{Experimental Setup}
We designed an experimental setup (Fig.~\ref{fig1:exp}) to measure the reaction force $F$ of the facia lata in response to an elongation $L$. The experiment consists of a stretching experiment directly followed by a relaxation experiment.

The participant is positioned on a bench as shown in Fig.~\ref{fig1:exp}. Proper stabilization of the skeletal structure is essential~\cite{germain2020}. Specifically, fixation of both the pelvis and the contra lateral leg is required to maintain stability of the hip of the stretched leg in all three anatomical planes. To achieve this, the contra lateral leg was immobilized in extension using a sling positioned around the knee. The ankle joint angle was fixed, and the distance between the hip and the heel was kept constant using a bench that stabilized the contra lateral foot.
To further enhance the stability of the participants, the backrest was extended and lowered compared to ~\cite{germain2020} to prevent posterior pelvic displacement. The underlying principle was to use the participant’s rigid skeletal structure to fix the pelvis in a specific and repeatable position. Once the pelvis was secured, the stretched leg could be mobilized. For this purpose, a sling was used to maintain the ankle in the sagittal plane and a second customized sling was applied to pull the stretched leg posteriorly, thus inducing hip extension.
An electrical actuator (Lunak, Sweden; 2 cm/s) was incorporated into the system to generate hip extension and provide improved precision in regulating fascia lata lengthening speed.
A static 5 mm width dynema rope was used and the fascia lata lengthening was proportional to the travel of the rope: 5~mm of rope travel induces a 1 mm fascia lata lengthening. This approach was chosen because it was the best way to capture the reaction force of the fascia lata using the center axes of the hip joint without introducing the added friction of an exoskeleton. The reaction force ($F$) was measured on a lever arm 30 cm from the articular center of the hip joint. Technically, the lever arm used, magnifies the fascia lata lengthening by a factor 5. The force measurements were acquired with a PCE-DFG N~1K dynamometer (PCE Instruments, France) and recorded via the accompanying PCE software. The setup was calibrated to ensure that a lengthening rate of 0.125\%\,s$^{-1}$ produced measurable force changes. The stretch speed was controlled by the actuator and the participant could control the stretch magnitude with two buttons (+ to stretch more, - to release the stretch). The surface electrocardiograms (EMGs) of the rectus femoris and of the tensor fascia lata were simultaneously monitored using the Phenix 4 system (Vivaltis, Montpellier, France). EMG signals exceeding 2~$\mu$V, indicating active contraction, led to the exclusion of the corresponding data.

\begin{figure}
    \centering
    \includegraphics[width=0.45\textwidth]{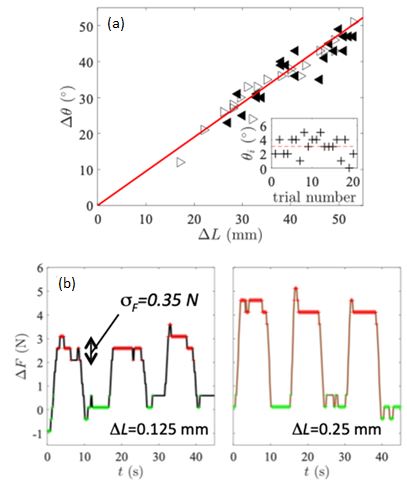} 
    \caption{Experimental setup characteristics. (a) Calibration of the fascia lata elongation $L$ as a function of the variation in inclination angle $\Delta \theta$ relative to the initial angle. The red line represents a linear fit: $\Delta \theta = 0.95 \Delta L$. Inset: Initial angle ($\theta_i$) for different trials to evaluate experiment reproducibility. The average value is $\langle \theta_i \rangle = 3.0 \pm 1.4^\circ$. (b) Measurement of the force variation $\Delta F$ as a function of time for repetitive elongations of the fascia lata from 0~mm to 0.125~mm (right) and 0.25~mm (left). The standard deviation of the plateau at $\Delta L = 0$ and 0.125 or 0.25~mm is independent of $\Delta L$, with a value of its standard deviation as $\sigma_F = 0.35$~N, setting the sensitivity of the force measurement to $2\sigma_F = 0.7$~N.}
    \label{fig2:callib}
\end{figure}

\subsection{Experiment limitations}
Among the limitations of the experiments, achieving reproducible pelvis fixation was challenging. To mitigate this, bony landmarks were used for the participant, resulting in a residual error margin of approximately $\pm 1.5^\circ$ for the hip extension angle $\theta_i$ as shown in Fig.~\ref{fig2:callib}(a).

To assess the sensitivity of our setup, we measured the force variation $\Delta F$ as a function of time during repeated elongations of the fascia lata from 0~mm to 0.125~mm and 0.25~mm, as shown in Fig.~\ref{fig2:callib}(b). Based on the standard deviation of the force relative to the plateau values at $L =0$,  0.125 and 0.25~mm, we estimate the experiment sensitivity to be of about 0.7~N.

Involuntary movements of participants, such as slight head or arm motions, occasionally compromised the recordings. Each experiment was therefore repeated multiple times to ensure clean data. Additionally, minor fluctuations in rectus femoris contractile activity during muscle rest were observed, affecting primarily the relaxation measurements. Overall, monitoring abnormal tension increases proved more effective for detecting unintended movements than relying solely on rectus femoris surface electromyography, which cannot capture involuntary movements occurring in other body regions.

\subsection{Rope properties}
\begin{figure}
    \centering
    \includegraphics[width=0.49\textwidth]{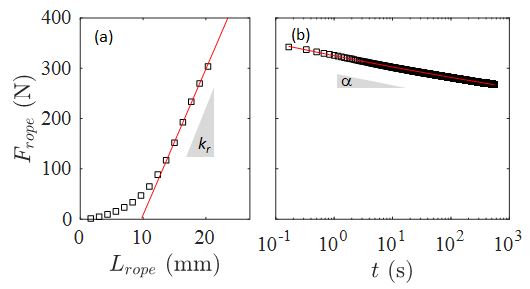} 
    \caption{Rope calibration: (a) extension and (b) relaxation.}
    \label{fig2:rope}
\end{figure}

Ideally, the rope would transmit force directly and behave as a perfectly rigid body. In practice, it is stiffer than the fascia lata but still exhibits a small, measurable compliance and a slow relaxation. Let us examine the stretching and relaxation of the rope alone. To do so, the length of rope used for the experiment is attached to a fixed 8 mm bolt. In the stretching test the rope is elongated at a speed $v=8~$mm/s. As shown in Fig.~\ref{fig2:rope}(a), the rope does not immediately follow $F_{\text{rope}} = k_r L_{\text{rope}}$, as the initial stretching first straightens kinks, aligns fibers, and overcomes internal friction. Only after this phase does the intrinsic elasticity dominate, with $k_r = 35~\text{N/mm}$. During the relaxation test, carried out at the end of the stretching step, the rope length $L$ is maintained constant. As shown in Fig.~\ref{fig2:rope}(b), we observe that the force slowly decays as 
$F_{\text{rope}}(t) \propto t^{-\alpha}$ with $\alpha = 0.031$.

Rather than neglecting the rope contribution, the rope was treated as an independent mechanical element in series with the fascia lata such that:
\begin{equation}
\left\{
\begin{aligned}
F_{\text{total}} &= F_{\text{fascia}} = F_{\text{rope}}, \\
L_{\text{total}} &= L_{\text{fascia}} + L_{\text{rope}}.
\end{aligned}
\right.
\end{equation}

\subsection{Numerical Solution of the Ramp–Relaxation experiment}

For the fascia stretching test, using data from Fig.~\ref{fig2:rope}(a), the fascia elongation was computed as $L_{\text{fascia}}=L_{\text{total}}-L_{\text{rope}}(F_{\text{total}})$, allowing the fascia response to be corrected for the rope compliance without assuming a specific constitutive model.

During the relaxation phase, correcting the relaxation test for the rope contribution is more challenging. Because the total length is held constant, $L_{\text{total}} = L_{\text{fascia}}(t) + L_{\text{rope}}(t) = \text{cst}$ any decrease in rope tension results in a corresponding increase in fascia elongation. Consequently, the system can only be solved using a constitutive model for the fascia lata.
%
In the relaxation test, the rope is represented as a linear elastic spring of fixed stiffness $k_r$, calibrated independently from rope-only tests, together with a slow power-law relaxation term:
\begin{equation}
L_{\mathrm{rope}}(t)
  = \frac{F(t)}{k_r}
    + L_{r0}\, (t/t_{r0})^{-\alpha},
\end{equation}
Evaluating this expression at the reference time $t = t_{r0} = 0.16~\mathrm{s}$ (the first point of the relaxation test), $L_{r0}$ represents the additional elongation associated with the power-law decay. In practice, $L_{r0}$ is obtained as the difference between the total rope elongation and its instantaneous elastic response, computed from the measured initial force $F(t_{r0})$ and the calibrated stiffness $k_r$: $L_{r0} = L_{\mathrm{rope}}(t_{r0}) - \frac{F(t_{r0})}{k_r}$.
This provides a direct measure of the amplitude of the slow relaxation component of the rope. This formulation captures the experimentally observed slow viscous relaxation of the rope while keeping its elastic response consistent with the calibration data. We then compute the fascia force $F(t)$  during relaxation using the intrinsic constitutive model as developed in section III.b with a fixed-point iteration to enforce consistency with the series constraint. An interactive implementation of the ramp--relaxation model is provided in the appendix as a supplementary material.
Finaly, a nonlinear least-squares minimization is applied to fit the experimental data. The rope is modelled in series with the fascia lata constitutive model, so that the minimization yields the effective facia lata parameters independently of the rope compliance.

\section{Results}

\subsection{Experiment: stretching and relaxation}

\begin{figure}
    \centering
    \includegraphics[width=0.45\textwidth]{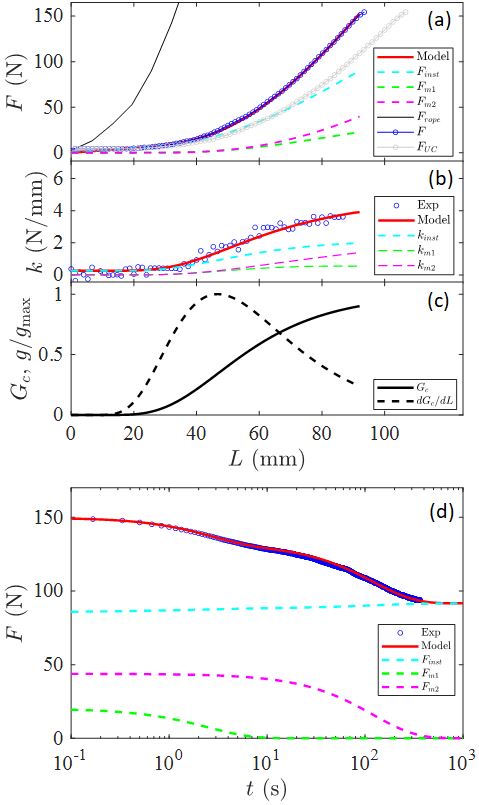} 
\caption{Stretching--relaxation experiment. (a) Stretching experiment. The grey dots represent the raw fascia data, the black line the contribution of the rope, and the blue dots the fascia data corrected for the rope contribution. The red line corresponds to the fit of the fascia lata response using the model. The dashed lines represent the different contributions of the model to $F$. 
(b) Stiffness. 
(c) Recruitment functions. 
(d) Relaxation experiment. Note that in the $F(t)$ representation there is a unique curve because, in series, $F_{\text{total}} = F_{\text{fascia}} = F_{\text{rope}}$.
The fitted parameters are $k_m = 0.26\,\mathrm{N\,mm^{-1}}$, $k_f = 1.91\,\mathrm{N\,mm^{-1}}$, $L_0 = 55\,\mathrm{mm}$, $\sigma = 0.4$, $k_1 = 0.39\,\mathrm{N\,mm^{-1}}$, $\tau_1 = 2.16\,\mathrm{s}$, $k_2 = 19.0\,\mathrm{N\,mm^{-1}}$, and $\tau_2 = 136\,\mathrm{s}$.
}
    \label{fig4:raw}
\end{figure}

The experiment consists of two consecutive steps: a stretching phase (Fig.~\ref{fig4:raw}(a)) immediately followed by a relaxation phase (Fig.~\ref{fig4:raw}(d)).
During stretching, the fascia lata of initial length $L$ is elongated at a constant speed of $v = 4$ or 8~mm/s up to a maximum force $F_0$ between 80 and 160~N. The relaxation phase then consists of maintaining the fascia lata at this constant length $L_{\text{max}}$ reached at $F_0$ during the elongation for 5 to 10~min. In both steps, the force is recorded at a sampling rate of 6~Hz. We have performed this experiment on a single subject and we have measured the response of left and right leg and repeated the experiment multiple times vayring $v$ and $F_0$.

A typical response of the fascia lata to elongation is shown in Fig.~\ref{fig4:raw}(a). The measured force $F$ increases slowly at first, then more steeply, as elongation $L$.
The corresponding stiffness $k = dF/dL$ exhibits a sigmoidal-like dependence on $L$, reflecting the progressive recruitment of spring-like elements during elongation (see Fig.~\ref{fig4:raw}(b).
During the relaxation test (Fig.~\ref{fig4:raw}(d), the force recorded at constant length decays in two distinct time steps, indicating a fast process characterized by $\tau_1$ and a slower one characterized by $\tau_2$.
In the literature, elongation and relaxation behaviors have generally been analyzed separately. Despite these methodological differences, our in vivo results show qualitative—and in some cases quantitative—agreement with prior experimental work on fascia lata and related soft connective systems in vivo and in vitro~\cite{szotek2024, otsuka2018, eng2014, kirilova2012, bonaldi2023, ristaniemi2021, shen2011, andriotis2023}.
Based on these observations, we propose a new model for the mechanical response of the fascia lata: a fiber–recruitment–viscoelastic framework incorporating strain–dependent memory effects.

\subsection{Model: fiber–recruitment–viscoelastic approach incorporating strain–dependent memory}
The total force developed by the fascia lata is decomposed as
\begin{equation}
F(t) = F_{\mathrm{inst}}(L) + F_{m,1}(t) + F_{m,2}(t),
\label{eq:total_force}
\end{equation}
where $F_{\mathrm{inst}}$ is the instantaneous elastic response and $F_{m,i}$ are two Maxwell-type viscoelastic contributions with distinct relaxation times.

\begin{figure}
    \centering
    \includegraphics[width=0.35\textwidth]{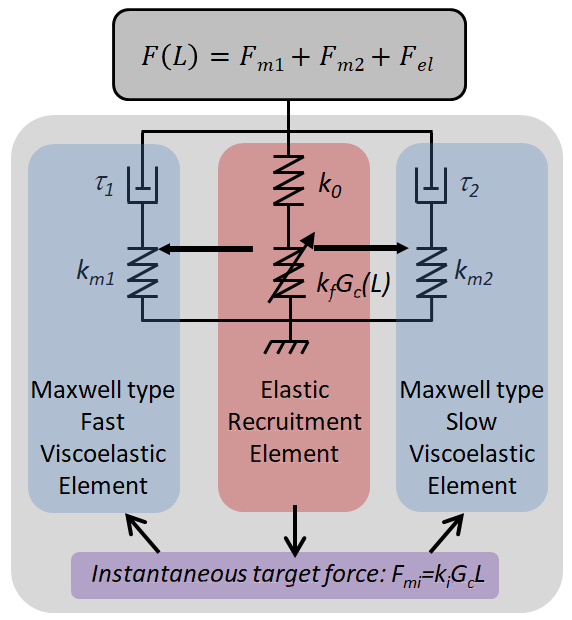} 
    \caption{Sketch of the fiber–recruitment–viscoelastic approach incorporating strain–dependent memory}
    \label{fig3:model}
\end{figure}

\vspace{2mm}
\paragraph{Elastic Contribution: Fiber Recruitment. }
The elastic force $F_{\mathrm{inst}}(L)$ arises from progressive recruitment of collagen fibers as the tissue is stretched. The instantaneous stiffness is
\begin{equation}
k_{\mathrm{inst}}(L) = k_0 + k_f \, G_c(L),
\end{equation}
where $k_0$ is the baseline matrix stiffness, $k_f$ is the total recruited fiber stiffness, and $G_c(L)$ is the cumulative distribution function (CDF) of the slack-length distribution $g(s)$, interpreted as the fraction of fibers recruited at length $L$. 

A convenient and physically interpretable choice for $G_c(L)$ is the lognormal distribution, which also captures the progressive recruitment of fibers while allowing for asymmetry in the recruitment process:
\begin{equation}
G_c(L) = \frac{1}{2} \left[1 + \mathrm{erf}\left(\frac{\ln L - \mu}{\sqrt{2}\sigma}\right)\right],
\end{equation}
where $\mu$ and $\sigma$ are the mean and standard deviation of $\ln L$, respectively. Here, $\mu$ determines the characteristic (median) recruitment length $L_0 = e^{\mu}$, and $\sigma$ controls the spread of fiber recruitment.
The corresponding probability density function is
\begin{equation}
g(s) = \frac{1}{s \sigma\sqrt{2\pi}}
\exp\left[-\frac{(\ln s - \mu)^2}{2\sigma^2}\right].
\end{equation}

The corresponding instantaneous elastic force is then
\begin{equation}
F_{\mathrm{inst}}(L) = \int_0^L k_{\mathrm{inst}}(s) ds.
\label{eq:elastic_force}
\end{equation}

The elastic component $F_{\mathrm{inst}}$ represents the permanent, crosslinked collagen network that does not relax on experimental timescales. Once fibers are recruited (governed by $G_c(L)$), they provide a non-dissipative structural backbone.

\vspace{2mm}
\paragraph{Viscoelastic Contributions: Maxwell Branches. }
Dissipative behavior is represented by two Maxwell elements arranged in parallel. In each branch is composed of a linear elastic spring (stiffness $k_{m,i}$) in series with a viscous dashpot, with a viscosity corresponding to a relaxation time $\tau_i$. we choose $\tau_1<<\tau_2$ to reflect the multiscale nature of the tissue microstructure.
The fast branch (i=1) models interstitial fluid redistribution and local fibril sliding; the slow branch (i=2) accounts for global fiber reorganization and compartmental fluid exchange. Each branch evolves according to a first-order relaxation law:
\begin{equation}
\tau_i \frac{dF_{m,i}}{dt} + F_{m,i} = F_{m,i}^{\ast}(L), \quad i=1,2,
\label{eq:maxwell_ode}
\end{equation}
where the instantaneous target force is
\begin{equation}
F_{m,i}^{\ast}(L) = k_{m,i}\, G_c(L)\, L.
\end{equation}

\vspace{2mm}
\paragraph{Coupling between the fiber recruitment and the viscoelastic branches. }
The instantaneous target force $F_{m,i}^{\ast}$ reflects two fundamental physical principles. First, its linear dependence on $L$ captures the elastic response of the spring element within each Maxwell branch. Second, the recruitment factor $G_c(L)$ ensures that viscoelastic dissipation scales with the proportion of load-bearing fibers. Slack fibers contribute negligibly to viscous resistance, whereas recruited, taut fibers generate significant fiber–matrix drag and internal friction. Consequently, at small extensions ($L \ll L_0$), where $G_c \approx 0$, viscous forces remain minimal. As recruitment progresses ($G_c \to 1$ for $L \gg L_0$), the full viscoelastic response becomes active. 

The instantaneous target force $F_{m,i}^{\ast}$ provides a direct coupling between fiber recruitment and viscoelastic dissipation, which is essential for reproducing the strain-dependent viscosity observed in fibrous connective tissues. This formulation leads to mechanical memory in their time-dependent mechanical response.
During a ramp–relaxation experiment, the presence of mechanical memory profoundly affects the observed force response. As the fascia lata is stretched at a constant rate, the recruited fibers and viscous elements accumulate internal stress according to their relaxation time spectra. Once the elongation is held constant, the subsequent relaxation does not start from a purely elastic state but from a history-dependent configuration that reflects both the rate and duration of the preceding ramp. As a result, the relaxation curve depends not only on the final strain but also on the strain rate and the temporal sequence of loading. In the proposed model, this manifests as a slower or incomplete stress relaxation after faster ramps, since a larger fraction of Maxwell elements remain partially engaged. Such behavior evidences the material's intrinsic memory, a hallmark of the fascia lata’s hierarchical viscoelastic structure where both collagen fibril recruitment and fluid-mediated dissipation retain information from prior loading events.

\begin{figure*}
    \centering
    \includegraphics[width=1\textwidth]{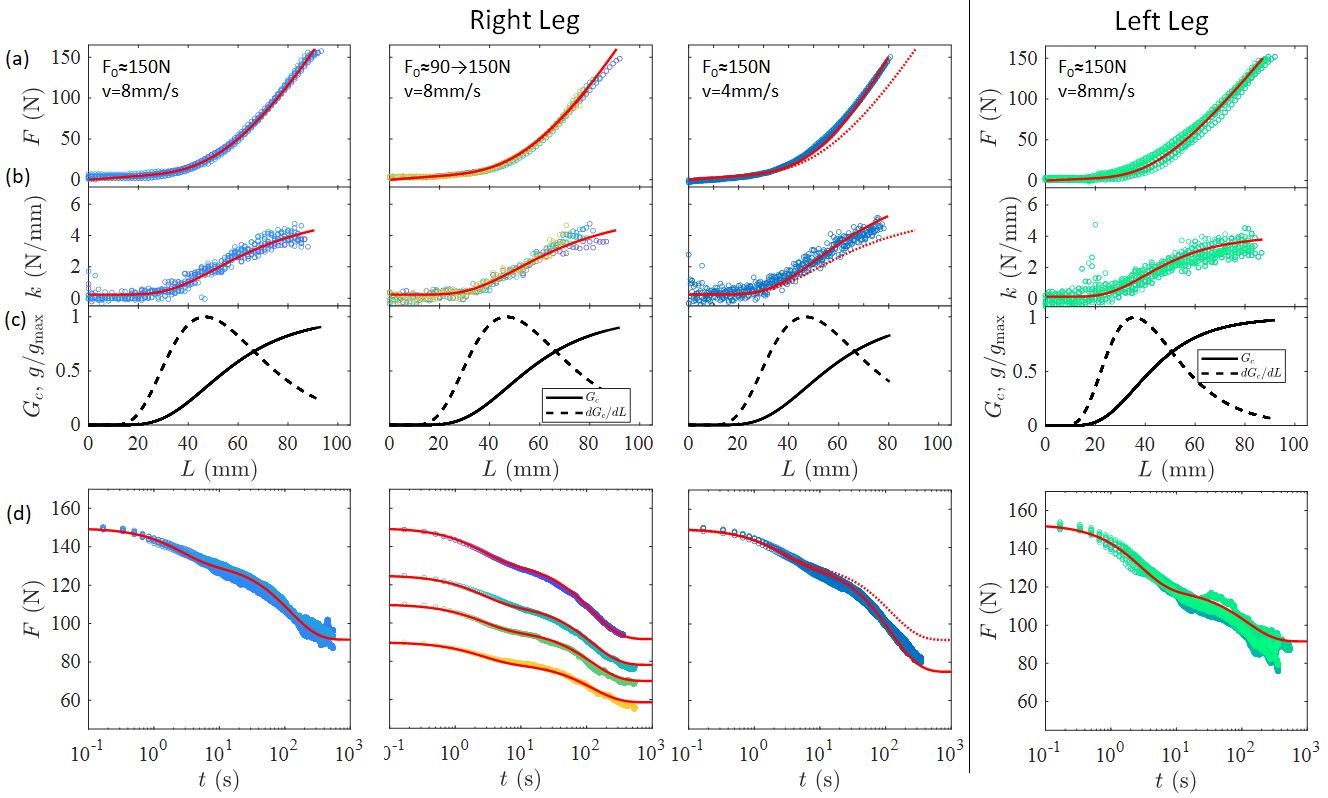} 
    \caption{Fascia lata experiment: right leg (first three panels) and left leg (fourth panel). (a--c) Stretching phase: force $F$, stiffness $k$, and recruitment function as functions of elongation $L$. (d) Force relaxation as a function of time. 
In the three right-leg panels, different conditions are shown. Panel 1: 10 measurements at ($F_0 \simeq 150$~N, $v = 8$~mm/s). Panel 2: four measurements at different $F_0$ with $v = 8$~mm/s. Panel 3: eight measurements at ($F_0 \simeq 150$~N, $v = 4$~mm/s). Panel 4 shows the results for the left leg at ($F_0 \simeq 150$~N, $v = 8$~mm/s). 
The solid red curves correspond to the model using average parameters over the entire series. The dashed red curves in panel 3 are shown for comparison and correspond to the model with $v = 8$~mm/s. The optimized recruitment distribution parameters yields ($\sigma = 0.4$, $L_0 = 55~\mathrm{mm}$) for the right leg and ($\sigma = 0.38$, $L_0 = 42~\mathrm{mm}$) for the left leg.
}
    \label{fig3:expG}
\end{figure*}

\begin{figure}
    \centering
    \includegraphics[width=0.48\textwidth]{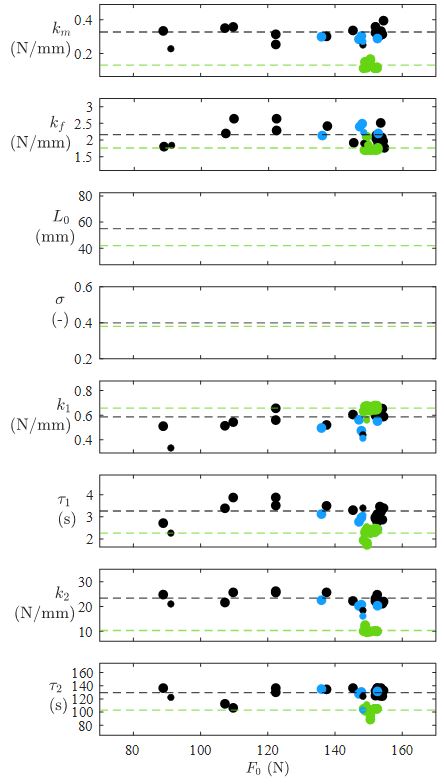} 
    \caption{Fascia lata fit parameters as function of $F_0$. right leg series with v=8mm/s (black dots), v=4mm/s (blue dots). left leg v=8mm/s (green dot). The dash lines represent the average value over the  right (black) and left (green) leg series. }
    \label{fig3:error}
\end{figure}

\subsection{Numerical Solution of the Ramp--Relaxation Experiment}

\paragraph{Numerical framework.}
The system consists of a rope in series with the fascia lata. The force $F$ is computed numerically by simulating two successive phases: a ramped elongation followed by a relaxation phase, according to the method described in Section~II.D. In Fig.~\ref{fig4:raw}, we display the raw experimental data together with the decomposition of the total response into rope and fascia lata contributions obtained from the fitting procedure.

\paragraph{Boundary condition during relaxation.}
During the relaxation phase (Fig.~\ref{fig4:raw}), the rope shortens as the force decreases from approximately 150 to 90~N, corresponding to a length change of $\Delta L_{\mathrm{rope}} = \Delta F / k_r \approx 60/35 \approx 1.7$~mm. Due to the series constraint ($L_{\mathrm{total}} = mathrm{const.}$), the fascia lata elongates by the same amount, representing approximately $2\%$ of its total elongation ($L_{\mathrm{max}} \approx 80$~mm). Although small, this indicates that the fascia does not undergo strictly constant-length relaxation, but instead experiences a slow residual elongation as the system equilibrates. At this strain level, most load-bearing fibers are already recruited ($G_c(L) \gtrsim 0.83$), so this additional stretch has only a limited effect on fiber recruitment and viscous dissipation mechanisms. The force decreases by approximately $40\%$, while the length increases by only $2\%$, indicating that viscoelastic dissipation dominates over elastic contributions induced by the boundary condition.

{The impact of this rope length drift during relaxation on the fitted relaxation times can be estimated by treating it as a small perturbation to an ideal constant-length relaxation. At large elongation, the recruitment function is nearly saturated, so the instantaneous stiffness is approximately constant, $k_{\mathrm{inst}} \approx k_\infty$. The additional elastic contribution therefore scales as $\delta F \sim k_\infty \Delta L$.
Using $\Delta L / L \approx 4\%$ and defining the relaxation amplitude as $A = F_0 - F_\infty$, with $F_0 \approx 160$~N and $F_\infty \approx 100$~N, one obtains $A/F_0 \approx 40\%$ and thus $\delta F / A \sim (\Delta L/L)/(A/F_0) \approx 0.1$. The resulting bias on the characteristic times scales as $\delta \tau / \tau \sim (\delta F / A)/\ln(F_0/F_\infty)$. With $\ln(160/100) \approx 0.47$, this yields $\delta \tau / \tau \sim 0.1$--$0.2$.
Because the slow increase in fascia length introduces an additional elastic contribution that partially compensates the force decay, the measured relaxation appears slightly slower than the intrinsic response, leading to a systematic overestimation of the fitted relaxation times. This estimate represents an upper bound, and the actual error is expected to be closer to $\sim 10\%$.}

\paragraph{Recruitment distribution.}
{We propose that fibril recruitment in the fascia lata is more appropriately described by a lognormal distribution rather than a sigmoidal one \cite{germain2016mecanismes,germain2020,germain2023stretch}. This behavior may be related to the geometrical characteristics of the fascial sheet. Unlike tendons, where the small cross-sectional area and nearly parallel fibril orientations result in largely uniform recruitment, fibril activation in fascia depends on the spatial extent of the tissue involved in the stretch. Fibrils located more laterally or along less favorable trajectories are less likely to be recruited, introducing heterogeneity in load sharing across the fascial surface. This variability, arising from multiple independent geometrical factors, naturally produces a skewed, multiplicative effect on recruitment, consistent with a lognormal distribution rather than a symmetric or uniform one.  
In Fig.~\ref{fig3:expG}(b), the stiffness $k$ exhibits a slight increase at large elongations $L$ rather than reaching a strict plateau. This behavior is consistent with the asymmetric, long-tailed character of the lognormal distribution, which allows for continued fiber recruitment at high $L$. Such a feature cannot be captured by symmetric or uniform distributions, which lack the ability to represent delayed recruitment in the high-elongation regime.}

{\paragraph{Fitting strategy.}
A brute-force fit applied independently to each measurements leads to large variability in the estimated parameters, with relative errors of about $40\%$. To improve the consistency of the fit across the data series, we constrain the recruitment distribution by keeping $L_0$ and $\sigma$ constant for all datasets. This choice is motivated by the fact that these parameters describe the structural distribution of fiber engagement and are therefore expected to remain constant across the series, as all measurements are performed on the same specimen. In contrast, the mechanical parameters (e.g., $k_0$, $k_f$, $k_1$, $k_2$, $\tau_1$, $\tau_2$) represent effective stiffness and viscoelastic responses, which are more sensitive to loading history and parameter correlations. Allowing these parameters to vary while constraining $L_0$ and $\sigma$ reduces parameter degeneracy and improves the robustness and consistency of the fits.}

\begin{figure}
    \centering
    \includegraphics[width=0.45\textwidth]{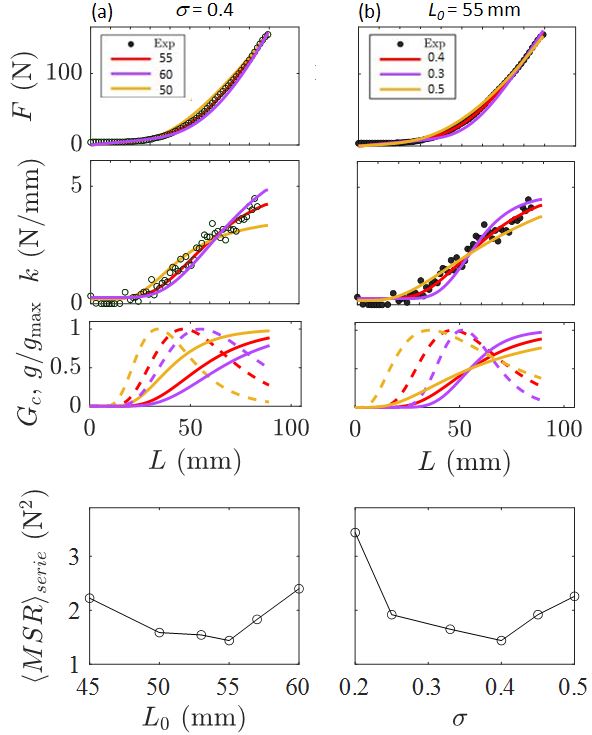}
    \caption{Optimisation of the recruitment distribution parameters ($\sigma$, $L_0$) for the right leg.
(a)~Influence of $L_0$ at $\sigma = 0.4$ on a representative acquisition
at ($v = 8$~mm/s, $F_0 \simeq 150$~N).
(b)~Influence of $\sigma$ at $L_0 = 55$~mm on the same dataset.
In both panels, rows from top to bottom show $F$, $k$, $G_c$, $g/g_{\max}$,
and the mean squared residual~($MSR=\langle \sum_{points}(F_{exp}-F_{model})^2\rangle_{series}$) as a function of $L_0$ or $\sigma$. Red curves correspond to the optimized value.
The MSR is evaluated over the full dataset of 17 independent measurements
of the right leg at $v = 8$~mm/s, with each individual fit performed jointly
on the ramp and relaxation phases. We restrict the comparison to the stretching phase, which is the most sensitive indicator of the recruitment distribution.}
    \label{fig4:msr}
\end{figure}

{The optimal values of $L_0$ and $\sigma$ are determined by systematically varying these parameters and minimizing the average mean square residual (MSR) over the series. As shown in Fig.~\ref{fig4:msr}, the optimal values are $L_0 = 55\,\mathrm{mm}$ and $\sigma = 0.4$ for the right leg and $L_0 = 42\,\mathrm{mm}$ and $\sigma = 0.38$ for the left leg. Using these optimal distribution parameters leads to an error of approximately $10\%$ on the mechanical parameters (Fig.~\ref{fig5:err}), providing an estimate of the intrinsic precision of the experimental setup.}


\paragraph{Local sensitivity and parameter identifiability.}
\tg{To assess the local influence of each free mechanical parameter on the objective function, we computed a normalized cost variation index~\cite{Aster2012, Saltelli2004, Ljung1999} for $\theta_i \in \{k_m, k_f, k_1, \tau_1, k_2, \tau_2\}$. Each parameter was perturbed by $\delta = \pm 5\%$ around its optimal value while all other parameters were held fixed. The normalized sensitivity index is defined as $S_i = \left| \frac{\mathrm{MSR}(\theta_i^+) - \mathrm{MSR}(\theta_i^-)}{\mathrm{MSR}_0} \right|,$ where $\theta_i^{\pm} = \theta_i (1 \pm \delta)$ and $\mathrm{MSR}_0$ denotes the objective value at the optimum. The resulting values indicate that a $\delta = \pm 5\%$ perturbation lids to relative chance in the $MSR$ of 40 to 180\%. All parameters influence the objective function locally, although with markedly different strengths.}

\begin{figure}
    \centering
    \includegraphics[width=0.48\textwidth]{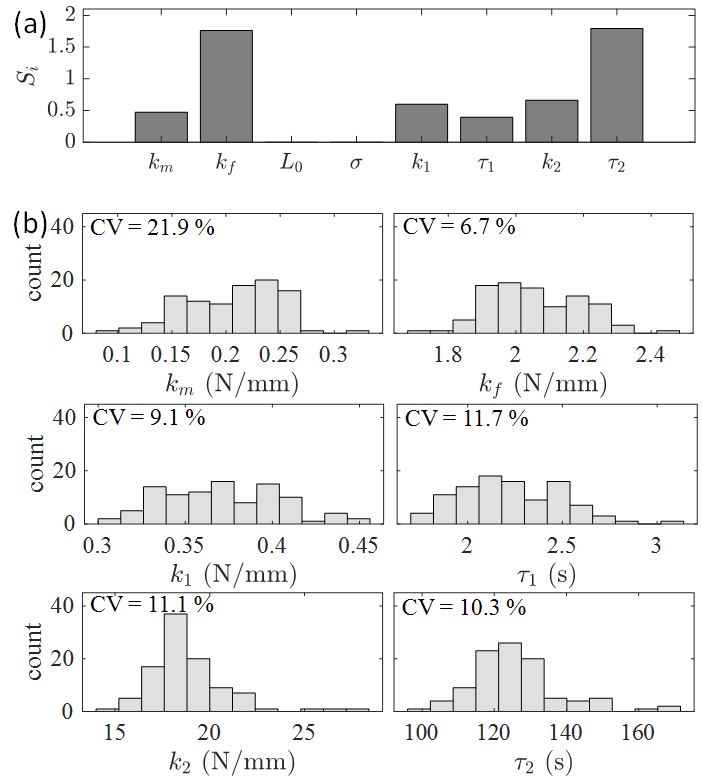}
   \caption{Representative sensitivity and uniqueness of the fitting procedure. 
(a)~Normalised sensitivity index $S_i$ for each free mechanical parameter0 
(b)~Distributions of fitted parameters across 100 optimisation runs 
from initial conditions drawn uniformly at random within $\pm50$\,\% 
of the optimal values; the coefficient of variation $\mathrm{CV}_i$ 
is indicated above each histogram.}
    \label{fig:scv}
\end{figure}

\tg{To assess the uniqueness of the identified minimum and the robustness of the optimisation landscape, a multi-start optimisation strategy was employed, in which the fitting procedure was repeated 100 times from random initial conditions drawn uniformly within ±50\% of the optimal parameter values. This approach is commonly used to evaluate the stability of nonlinear inverse problems and the dependence of the solution on initial conditions \cite{Aster2012, Nocedal2006, Horst1995}. For each run, the optimiser successfully converged back to the vicinity of the optimal parameter set $\boldsymbol{\theta_i}$. Parameter consistency across these runs is quantified by the coefficient of variation $\mathrm{CV}_i = 100\times\frac{\sigma_i}{\mu_i}\quad[\%]$, where $\mu_i$ and $\sigma_i$ are the mean and standard deviation of $\theta_i$ across the 100 runs. The resulting distributions are shown in Fig.~\ref{fig:scv}(b). All six parameters yield $\mathrm{CV}_i \in [7,\,22]\,$\%, indicating that repeated optimisations from widely dispersed initial conditions consistently converge to a narrow region in parameter space. This behaviour suggests the presence of a dominant and stable minimum in the objective landscape, with limited sensitivity of the final solution to initialisation. 
}

\vspace{2mm}
{\paragraph{An effective model.}
Finally, we emphasize that the measured force represents a global response of the loaded hip--thigh complex, while local stress and strain fields within the fascia lata are necessarily heterogeneous. The constitutive framework proposed here should therefore be interpreted as an effective one-dimensional macroscopic law under controlled boundary conditions, rather than a local three-dimensional stress--strain description. Under the imposed kinematics, hip extension predominantly loads the longitudinal axis of the fascia lata, allowing definition of a dominant loading direction and a corresponding effective elongation. The fitted parameters thus characterize the homogenized mechanical response of the fascia-dominant structure along this axis. Validation of spatial stress and strain distributions would require imaging-based strain mapping or multiscale modeling, which are beyond the scope of the present proof-of-concept study.}

\paragraph{Comparison with other models.}
The central modeling challenge in a ramp--relaxation experiment is that elastic and viscous contributions are not separable: viscoelastic elements accumulate internal stress during fiber recruitment, so that the relaxation response encodes the full loading history of the preceding ramp. Existing recruitment-based models \cite{romero1998recruitment, raz2009recruitment, bevan2018biomechanical} treat fiber engagement and viscous dissipation as independent additive contributions, and therefore cannot reproduce this history dependence or the resulting strain-dependent viscosity. Microstructural models are in principle capable of capturing these phenomena, but their parameterization exceeds what sparse in vivo data can support. Generalized \cite{stecco2009} or fractional Maxwell models \cite{stankiewicz2018} fit relaxation data accurately but are purely phenomenological and contain no recruitment mechanism.

Our approach occupies the gap between these extremes: by coupling each Maxwell branch to the recruitment function $G_c(L)$ multiplicatively, viscous dissipation scales naturally with the load-bearing fiber fraction, mechanical memory across the ramp--relaxation transition emerges without additional assumptions, and both experimental phases are described by a single, shared, physically interpretable parameter set --- a prerequisite for meaningful longitudinal tracking.

\subsection{Analysis of the experiments, fit results}

\begin{figure}
    \centering
    \includegraphics[width=0.475\textwidth]{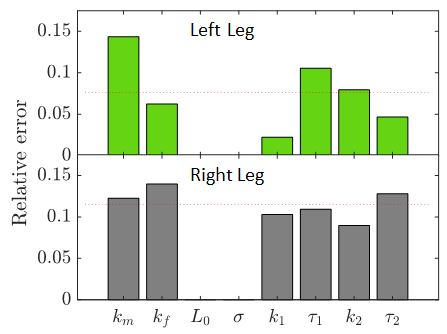} 
\caption{Relative error of the fit parameters obtained from individual fits performed on the left and right legs. The optimized recruitment distribution parameters were kept constant during the fitting process: for the right leg, $\sigma = 0.4$ and $L_0 = 55,\mathrm{mm}$; for the left leg, $\sigma = 0.38$ and $L_0 = 42,\mathrm{mm}$. The error on each parameters corresponds to the standard deviation based on the fit results in Fig.~\ref{fig3:error}. The red dot line correspond to the average relative error over all the parameters. 
}
    \label{fig5:err}
\end{figure}

\begin{table*}
\centering
\begin{tabular}{|l|c|c|c|p{8.5cm}|}
\hline
\textbf{Parameter} & \textbf{Symbol} & \textbf{Left} & \textbf{Right} & \textbf{ Mechanism (hypothetical)} \\
\hline
\multicolumn{5}{|l|}{\textit{Elastic fiber recruitment}} \\
\hline
Matrix stiffness & $k_0$ (N/mm) & 0.14 & 0.33 & GAG-mediated hydration and inter-fiber gliding provide baseline elastic resistance. \\
Fiber stiffness & $k_f$ (N/mm) & 1.85 & 2.15 & Collagen and elastin fibers bear load once recruited, setting high-strain stiffness. \\
Recruitment length & $L_0$ (mm) & 42 & 55 & Collagen crimp sets the onset of fiber engagement. \\
Recruitment width & $\sigma$ & 0.38 & 0.40 & Variation in fiber lengths controls progressive recruitment. \\
\hline
\multicolumn{5}{|l|}{\textit{Maxwell branches}} \\
\hline
Fast branch & 
\begin{tabular}{c}
$k_1$ (N/mm) \\
$\tau_1$ (s)
\end{tabular} & 
\begin{tabular}{c}
0.65 \\
2.1
\end{tabular} & 
\begin{tabular}{c}
0.65 \\
3.1
\end{tabular} & 
Short-term viscoelastic dissipation via collagen--GAG and fiber--matrix interactions. \\
\hline
Slow branch & 
\begin{tabular}{c}
$k_2$ (N/mm) \\
$\tau_2$ (s)
\end{tabular} & 
\begin{tabular}{c}
10.1 \\
102
\end{tabular} & 
\begin{tabular}{c}
20.5 \\
129
\end{tabular} & 
Long-term viscoelastic relaxation via fluid flow, osmotic effects, and extracellular matrix rearrangements. \\
\hline
\end{tabular}
\caption{Fitted model parameters averaged over the series for the fascia lata. Each parameter is associated with a plausible underlying physical mechanism (see section discussion).}
\label{tab:param}
\end{table*}

{In Fig.~\ref{fig4:raw} and Fig.~\ref{fig3:expG}, we present the results of the 
ramp--relaxation experiments and the corresponding model fits for all experimental 
series. The total force $F_{\mathrm{tot}}$ combines the elastic fiber contribution 
$F_{inst}$ with the fast and slow Maxwell responses $F_{m,1}$ and $F_{m,2}$ 
respectively. We applied this framework to four distinct experimental series, each 
probing a specific aspect of the fascia lata mechanical response. Panel 1 and 
Panel 4 of Fig.~\ref{fig3:expG} consist of repeated ramp--relaxation experiments 
performed on the right and left leg respectively, at fixed velocity $v = 8$~mm/s 
and maximum force $F_0 \simeq 150$~N, allowing the reproducibility of the fitted 
parameters to be assessed under controlled loading conditions. Panel 2 comprises 
right-leg experiments in which $F_0$ was systematically varied from approximately 
90 to 160~N at $v = 8$~mm/s, probing the dependence of the mechanical response on 
loading amplitude. Panel 3 examines the effect of reducing the loading velocity to 
$v = 4$~mm/s at fixed $F_0 \simeq 150$~N on the right leg, probing the influence 
of loading rate on the viscoelastic response. We have uploaded to Github the matlab code used to fit our experimental data under stretching-relaxation conditions. We have also provided experimental data to test the code~\cite{gibaud2026}. }

{\paragraph{Model fit and force decomposition: representative experiment 
(Fig.~\ref{fig4:raw}).}
Fig.~\ref{fig4:raw} shows a single representative trial on the right leg at 
$v = 8$~mm/s and $F_0 \simeq 150$~N, together with the full decomposition of the 
model force into its three contributions. The rope correction significantly affects 
the extension phase: it reduces the apparent fascia elongation $L$ and increases 
the local slope $\mathrm{d}F/\mathrm{d}L$. During the ramp, the dominant 
contributions to the total force are the build-up of the equilibrium force 
$F_{inst} = \int_0^L k_{\mathrm{inst}}(L')\,\mathrm{d}L'$, where 
$k_{\mathrm{inst}} = k_m + k_f G_c(L)$ is the instantaneous stiffness, and of the 
fast Maxwell force $F_{m,1}$. The rise of $F_{inst}$ is governed by the 
recruitment function $G_c(L)$, which increases sharply around $L_0$, producing 
the characteristic sigmoidal stiffening. Since the ramp duration (${\sim}10$~s) 
is comparable to the characteristic time of the fast Maxwell branch 
($\tau_1 \simeq 2$~s), $F_{m,1}$ has sufficient time to accumulate a significant 
load during stretching. In contrast, the slow Maxwell branch $F_{m,2}$ remains 
largely unloaded during the ramp, as its characteristic time ($\tau_2 \simeq 
130$~s) greatly exceeds the ramp duration. During the subsequent relaxation, the 
two Maxwell branches decay with their respective characteristic times $\tau_1$ and 
$\tau_2$, with initial amplitudes set by the values of $F_{m,1}$ and $F_{m,2}$ 
reached at the end of the extension phase. At long times, the force approaches 
$F_{inst}(L_{\mathrm{max}})$, the elastic backbone evaluated at the maximum 
elongation. The rope correction manifests during relaxation as a slight increase of 
$F_{inst}$ of approximately $4$~N over the relaxation duration, 
corresponding to an inelastic rope recovery of approximately $1$~mm that slowly 
transfers length from the rope to the fascia, marginally stretching it further and 
raising the equilibrium force along the recruitment curve.}

{\paragraph{Reproducibility: right leg at $v = 8$~mm/s and $F_0 \simeq 150$~N 
(Fig.~\ref{fig3:expG}, Panel 1).}
Panel 1 of Fig.~\ref{fig3:expG} shows the repeated trials at the reference 
conditions. The model reproduces both the stretching and relaxation phases 
simultaneously across all trials with small residuals, and the fitted parameters 
are recovered with a relative uncertainty of approximately $10\%$, which we take 
as the intrinsic sensitivity of the experimental protocol. This variability has two 
distinct origins. The first is instrumental: despite the use of bony landmarks to 
ensure repeatable positioning, the hip extension angle carries a residual 
uncertainty of $\pm 1.5^\circ$ once the pelvis is secured, and slight variations 
in rope travel arising from initial positioning introduce additional scatter in the 
imposed elongation. The second origin is physiological, and is likely the dominant 
contribution. Stretch tolerance is known to increase significantly during a 
sustained 10-minute stretch~\cite{germain2016mecanismes}, and repeated stretching 
modifies both stretch tolerance and fibril recruitment~\cite{germain2020}. As the 
fascia lata adapts progressively to warm-up and cumulative mechanical loading, its 
initial state at the onset of each trial cannot be perfectly controlled, 
constituting an irreducible lower bound on the reproducibility of any in vivo 
fascial measurement protocol.}

{\paragraph{Influence of loading amplitude: right leg at $v = 8$~mm/s and $F_0 = 
90$--160~N (Fig.~\ref{fig3:expG}, Panel 2).}
Panel 2 shows the effect of varying $F_0$ from approximately 90 to 160~N at fixed 
velocity $v = 8$~mm/s. During the stretching phase, all force--elongation curves 
follow the same trajectory regardless of $F_0$, confirming that the elastic and 
recruitment parameters are intrinsic to the loading path and insensitive to the 
amplitude at which the ramp is stopped. During the relaxation phase, two consistent 
observations emerge. First, the timescales $\tau_1$ and $\tau_2$ remain unchanged 
across the full range of loading amplitudes, confirming that the relaxation dynamics 
are governed by intrinsic material timescales of the fascia lata complex independent 
of the maximum force. Second, the long-time asymptote decreases systematically as 
$F_0$ decreases. This is a direct geometric consequence of the model: a lower $F_0$ 
is reached by stopping the ramp at a smaller $L_{\mathrm{max}}$, so the elastic 
backbone $F_{inst}(L_{\mathrm{max}})$ --- which sets the long-time 
relaxation asymptote --- is evaluated at a less recruited state and therefore 
yields a lower asymptotic force. The model reproduces both observations 
simultaneously across all values of $F_0$, demonstrating 
the robustness and internal consistency of the fitted parameter set over a broad 
range of loading conditions.}

{\paragraph{Influence of stretching velocity: right leg at $v = 4$~mm/s and $F_0 
\simeq 150$~N (Fig.~\ref{fig3:expG}, Panel 3).}
Panel 3 examines the effect of reducing the stretching velocity to $v = 4$~mm/s 
at fixed $F_0 \simeq 150$~N. Compared to the reference experiment at $v = 8$~mm/s, 
the force rises more steeply with elongation $L$ and reaches $F_0$ at a shorter 
total elongation $L_{\mathrm{max}}$. At lower velocity, the ramp duration doubles 
to approximately 20~s, giving the fast Maxwell branch sufficient time to partially 
relax \textit{during} the stretch itself. Its viscous contribution is therefore 
reduced and the measured response lies closer to the purely elastic backbone 
$F_{inst}$, which is steeper and reaches a given force level at smaller 
elongation. During the subsequent relaxation phase, the timescales $\tau_1$ and 
$\tau_2$ are consistent with those of the reference experiment, confirming that 
the relaxation dynamics are rate-independent once the final elongation is reached. 
The long-time asymptote is, however, lower than in the reference experiment, as a 
direct geometric consequence of the shorter $L_{\mathrm{max}}$: the elastic 
baseline $F_{inst}(L_{\mathrm{max}})$ is evaluated at a less recruited 
state and therefore yields a lower asymptotic force. Taken together, these 
observations are a direct manifestation of the mechanical memory encoded in the 
model, and illustrate why a controlled and reproducible stretching velocity is 
essential for meaningful comparisons between experimental sessions.}

{\paragraph{Left leg at $v = 8$~mm/s and $F_0 \simeq 150$~N 
(Fig.~\ref{fig3:expG}, Panel 4).}
Panel 4 presents the results for the left leg under identical protocol conditions 
to Panel 1. The overall structure of the response is qualitatively similar: the 
same sigmoidal stiffening during stretching and the same double-exponential 
relaxation are observed, and the model reproduces both phases with comparable 
accuracy. Quantitatively, however, systematic differences in the fitted parameters 
are evident. The left leg exhibits a shorter median slack length ($L_0 = 42$~mm 
versus 55~mm), a lower fiber stiffness ($k_f = 1.85$ versus 2.15~N/mm), a reduced 
slow branch stiffness ($k_2 = 10.1$ versus 20.5~N/mm), and a slightly shorter slow 
relaxation time ($\tau_2 = 102$ versus 129~s), while the fast branch stiffness 
$k_1 = 0.65$~N/mm is identical between legs. These differences systematically 
exceed the $10\%$ experimental sensitivity and therefore reflect genuine mechanical 
asymmetry between the two limbs, attributable to the combined effects of limb 
dominance and the left leg's history of localized fascial injury, as discussed in the discussion
Section.}

{\paragraph{Fitted parameters and experimental sensitivity.}
The fitting procedure yields optimal recruitment distribution parameters of 
$L_0 = 55$~mm and $\sigma = 0.40$ for the right leg, and $L_0 = 42$~mm and 
$\sigma = 0.38$ for the left leg. These structural parameters are held fixed across 
trials within each series, as they describe the fiber slack-length distribution and 
are not expected to vary between successive measurements on the same subject. The 
remaining mechanical parameters --- $k_m$, $k_f$, $k_1$, $\tau_1$, $k_2$, 
$\tau_2$ --- are fitted independently for each trial and are displayed in 
Fig.~\ref{fig3:error}, where their variability across repeated trials is 
quantified. Across all series, these parameters are recovered with a relative 
uncertainty of approximately $10\%$ (Fig.~\ref{fig5:err}), which we interpret as the intrinsic sensitivity 
of the present experimental protocol and the baseline against which any 
training-induced or injury-related changes in fascial mechanics should be assessed.}
\subsection{Discussion}

\paragraph{Nonlinear stiffening and fiber recruitment.}
The stiffness of the fascia lata is not constant: the measured force--elongation 
response exhibits a characteristic nonlinear stiffening, consistent with similar 
behaviors observed both \textit{in vivo}~\cite{bennett1989} and \textit{in vitro} 
in rat-tail tendons~\cite{puxkandl2002}. Synchrotron-based studies on 
tendons~\cite{puxkandl2002} have shown that this response arises from progressive 
fibril alignment and load transfer from the hydrated matrix to collagen fibrils: 
the initial low-stiffness region corresponds to matrix deformation and slack fiber 
extension, while the subsequent steep rise reflects sequential fiber recruitment and 
full fibril engagement. This scenario is consistent with the predictions of our 
fiber--recruitment--viscoelastic model.

{\paragraph{Effective model and scope of parameter interpretation.}
Before interpreting the fitted parameters, it is important to recall that the 
constitutive framework proposed here is an effective one. The measured force 
reflects the global mechanical response of the fascia lata complex including its 
coupling to the hip--thigh musculoskeletal system, rather than a purely intrinsic 
local material property of the fascia lata itself. The fitted parameters therefore 
characterize the homogenized response of the fascia-dominant structure along the 
loading axis, under the specific boundary conditions of the ramp--relaxation 
protocol. Any interpretation of these parameters in terms of local stress, local 
strain, or nanoscale mechanisms should therefore be regarded as hypothetical, since 
we do not have access to local measurements. With this caveat in mind, the following 
paragraphs examine the physical plausibility of the fitted parameters in two 
complementary ways: by verifying that their values are of the same order of 
magnitude as those reported in the literature for comparable tissues, and by 
proposing tentative associations between macroscopic parameters and microstructural 
mechanisms that are consistent with, but not proven by, the present measurements. 
These hypothetical associations are certainly speculative, but by identifying which 
local mechanisms are most plausibly encoded in each macroscopic parameter, they 
lay the groundwork for future experiments combining the present protocol with local 
measurement techniques --- such as ultrasound strain imaging, fibril-scale 
diffraction, or multiphoton microscopy --- that could directly test and refine these 
interpretations.}

\paragraph{Elastic parameters: order of magnitude and hypothetical interpretation.}
The baseline matrix stiffness ranges from $k_0 = 0.14$~N/mm (left leg) to 
$k_0 = 0.33$~N/mm (right leg), reflecting in both cases a relatively compliant 
response consistent with the low resistance expected from a hydrated GAG-rich 
matrix with slack collagen at small strains. Assuming a typical fascia lata 
thickness of 0.3--0.5~mm in the longitudinal fiber layers of 
interest~\cite{otsuka2018,stecco2009,otsuka2021investigation} and a width of 
approximately 50~mm in the targeted stretch 
area~\cite{germain2023stretch,germain2024mechanical}, the estimated cross-sectional 
area is $A \approx 15$--25~mm$^2$. This yields an effective elastic modulus 
$E_0 = k_0 L / A \approx 0.39$--1.54~MPa (using $L \approx 70$~mm after extension 
and the full range of $k_0$ across both legs), consistent with the low-strain 
modulus of hydrated soft connective tissues~\cite{stecco2009}. The fully recruited 
fiber stiffness ranges from $k_f = 1.85$~N/mm (left leg) to $k_f = 2.15$~N/mm 
(right leg), representing the high-strain elastic response when collagen fibers are 
aligned and load-bearing. The total elastic stiffness at maximum recruitment, 
$k_0 + k_f \approx 1.99$--2.48~N/mm, corresponds to a modulus of approximately 
$E_f \approx 5.6$--11.6~MPa, in reasonable agreement with uniaxial tensile tests 
on human fascia lata reporting tangent moduli in the range 5--15~MPa at high 
strains~\cite{bonaldi2023}. The slightly lower values obtained here are consistent 
with the fact that our in vivo measurements capture the most compliant loading 
direction and include contributions from surrounding tissue compliance. These 
order-of-magnitude agreements lend plausibility to the following hypothetical 
interpretations: $k_0$ would most plausibly reflect GAG-mediated hydration pressure 
and inter-fiber gliding resistance at small strains, while $k_f$ would correspond 
to the collective stiffness of recruited, load-bearing collagen and elastin fibers 
at large strains. We emphasize, however, that these associations cannot be verified 
without local measurements such as fibril-scale strain mapping or imaging-based 
displacement fields.

\paragraph{Viscoelastic parameters: order of magnitude and hypothetical 
interpretation.}
The fast Maxwell branch is characterized by an identical stiffness $k_1 = 
0.65$~N/mm for both legs, and relaxation times of $\tau_1 = 2.1$~s (left leg) 
and $\tau_1 = 3.1$~s (right leg). The magnitude of $k_1$ is comparable to both 
$k_0$ and $k_f$, indicating that rapid viscous dissipation plays a substantial 
role in the immediate mechanical response, contributing approximately 30--40\% of 
the total force during the stretching phase. The associated dashpot coefficient 
$c_1 = k_1 \tau_1$ ranges from $1.4$~N$\cdot$s/mm (left leg) to 
$2.0$~N$\cdot$s/mm (right leg), corresponding to an effective viscosity 
$\eta_1 = c_1 L / A \approx 3.8$--9.4~MPa$\cdot$s (using $L \approx 70$~mm and 
$A \approx 15$--25~mm$^2$), of the same order of magnitude as values reported for 
hydrated collagenous tissues undergoing rapid fiber--matrix sliding and local water 
redistribution~\cite{gupta2009,screen2011,shen2011}. The slow Maxwell branch 
exhibits a substantially larger stiffness --- $k_2 = 10.1$~N/mm (left leg) and 
$k_2 = 20.5$~N/mm (right leg) --- and longer relaxation times of $\tau_2 = 102$~s 
and $\tau_2 = 129$~s respectively. The corresponding dashpot coefficients 
$c_2 = k_2 \tau_2 \approx 1.0$--2.6~kN$\cdot$s/mm yield effective viscosities 
$\eta_2 \approx 2.9$--12.3~GPa$\cdot$s, consistent with values reported for slow 
osmotic fluid redistribution through dense proteoglycan networks and large-scale 
fibrillar rearrangement in similar tissues~\cite{gupta2009,screen2011}. Although 
this branch is only partially activated during the 10~s loading ramp, it dominates 
long-term force relaxation. Notably, the identical $k_1$ across both legs contrasts 
with the marked left--right asymmetry observed in $k_2$, suggesting that the fast 
dissipative mechanism is robust to both limb dominance and injury history, while 
the slow branch is more sensitive to tissue remodeling. These order-of-magnitude 
agreements with the literature support, but do not prove, the following hypothetical 
associations: $\tau_1$ and $k_1$ most plausibly reflect fast collagen--GAG bond 
breaking and short-range water redistribution at the fibril scale, while $\tau_2$ 
and $k_2$ would correspond to slow fibrillar gliding, proteoglycan deformation, and 
large-scale osmotic fluid flow within the extracellular matrix. We stress that 
these interpretations go beyond what the macroscopic measurements alone can 
establish, and would require fibril- or matrix-scale experimental validation.

\paragraph{Comparison with the literature across scales.}
The measured relaxation timescales --- $\tau_1 = 2.1$--3.1~s and $\tau_2 = 
102$--129~s across both legs --- are consistent with a broad body of \textit{in 
vitro} literature spanning multiple scales, which serves the dual purpose of 
confirming that our fitted values are of the expected order of magnitude and of 
reinforcing the plausibility of the hypothetical microstructural interpretations 
proposed above. At the tissue scale, Duenwald \textit{et al.}~\cite{duenwald2010} 
reported dual relaxation in tendons with an initial timescale of approximately 10~s 
and a slower recovery of approximately 20~s, while Bonaldi \textit{et 
al.}~\cite{bonaldi2023} observed $\tau_1$ in the range 6--10~s and $\tau_2$ from 
approximately 124 to 200~s in frozen fascia lata, suggesting that certain 
viscoelastic behaviors are preserved despite sample freezing and ex vivo conditions. 
At the fibrillar scale, Screen \textit{et al.}~\cite{screen2011} and Gupta 
\textit{et al.}~\cite{gupta2009,shen2011} reported $\tau_1 \approx 4.5$--9.7~s 
and $\tau_2 \approx 82$--110~s in isolated collagen fibrils, attributing $\tau_1$ 
to water movement between fibrils and peripheral GAGs, and $\tau_2$ to fibrillar 
gliding or matrix rearrangement. At the molecular scale, Fullerton and 
Rahal~\cite{fullerton2007} showed that hydrogen-bonded water layers around collagen 
fibrils contribute to energy dissipation when released under load.
Our measured $\tau_1$ values (2.1--3.1~s) are somewhat shorter than most reported 
values for isolated fibrils or ex vivo tissue (4.5--10~s). This difference may 
reflect in vivo conditions that facilitate faster relaxation: higher physiological 
temperature (37$^\circ$C versus room temperature), active tissue perfusion, and 
reduced boundary constraints. Our measured $\tau_2$ values (102--129~s) fall within 
the range reported for isolated collagen fibrils (82--110~s) at the lower end and 
are somewhat shorter than those observed in frozen fascia lata (124--200~s), 
suggesting that the slow relaxation mechanism is an intrinsic property of the 
collagen--GAG composite that is relatively preserved across experimental conditions. 
The slight overlap with the upper end of the fibrillar range and the lower end of 
the frozen tissue range is consistent with in vivo conditions lying between these 
two extremes. When compared to muscle tissue, fascia lata relaxation is markedly 
faster: Gras \textit{et al.}~\cite{gras2013} modeled the human 
sternocleidomastoideus with $\tau_1 = 18$~s and $\tau_2 = 395$~s, likely 
reflecting the slower bound-water release associated with contractile 
proteins~\cite{matsuo2016}. This contrast reinforces that the fascia lata behaves 
mechanically as a dense collagenous tissue rather than muscle. Taken together, 
these multi-scale comparisons support, albeit indirectly, a two-process picture of 
fascial energy dissipation: rapid local dissipation from collagen--GAG interactions 
($\tau_1$) and slower large-scale structural reorganization within the extracellular 
matrix ($\tau_2$). A tentative mapping of each model parameter to its hypothetical 
molecular and structural origin is summarized in Table~\ref{tab:param}.

{\paragraph{Left--right asymmetry: dominance and injury history.}
The fitted parameters reveal a systematic mechanical asymmetry between the two 
limbs that exceeds the $10\%$ experimental sensitivity established above, and 
therefore reflects genuine differences in the effective mechanical response of the 
fascia lata complex rather than measurement scatter. The most salient differences 
concern the toe-region length $L_0$ and the long-term viscoelastic stiffness $k_2$: 
the right (dominant) leg exhibits $L_0 = 55$~mm and $k_2 = 20.5$~N/mm, against 
$L_0 = 42$~mm and $k_2 = 10.1$~N/mm for the left (see Tab.~\ref{tab:param}). 
Because the two legs differ simultaneously in dominance and injury history, their 
respective contributions cannot be disentangled from a single-subject dataset. The 
direction of the differences is nonetheless physically interpretable. The larger 
$L_0$ of the dominant leg reflects a fascial architecture remodeled toward greater 
extensibility under habitual mechanical loading; the shorter $L_0$ of the injured 
leg is consistent with a less-crimped, more disorganized collagen arrangement 
following post-traumatic repair. The twofold difference in $k_2$ points in the same 
direction: reduced slow-relaxation resistance in the injured leg is consistent with 
altered GAG content and collagen organization after scar 
remodeling~\cite{bonaldi2023,stecco2009}. The preservation of $k_1$ across legs 
suggests that the short-range dissipative mechanism is a robust intrinsic property 
of the collagen--GAG composite, relatively insensitive to remodeling. That the 
protocol resolves between-leg differences in $L_0$ and $k_2$ consistent with known 
adaptation and repair processes supports its potential as a quantitative tool in 
rehabilitation follow-up, provided dominance asymmetry is accounted for through 
contralateral control measurements.}

\paragraph{Physiological implications of effective stiffness and rate dependence.}
At the end of the stretching phase, the instantaneous stiffness $k_{\mathrm{inst}}$ 
is dominated by the recruited elastic fibers ($k_0 + k_f = 1.99$--2.48~N/mm, 
ranging from left to right leg) and the partially engaged fast Maxwell branch 
($k_1 = 0.65$~N/mm), yielding a total instantaneous stiffness of approximately 
2.6--3.1~N/mm. During steady-state loading at long times, the effective stiffness 
approaches $k_0 + k_f \approx 1.99$--2.48~N/mm as both Maxwell branches relax 
completely. However, during dynamic movements with intermediate loading timescales 
($\sim$1--10~s, typical of gait or jumping), both Maxwell branches contribute 
significantly --- the fast branch with $\tau_1 = 2.1$--3.1~s and the slow branch 
with $\tau_2 = 102$--129~s --- and the apparent stiffness can be 2--4 times higher 
than the purely elastic baseline due to rate-dependent viscous resistance. This 
strain-rate sensitivity is functionally important: it allows the fascia lata to 
provide elastic energy storage during slow, controlled movements and increased 
damping during rapid, high-impact activities. In terms of tissue modulus, using 
$A \approx 15$--25~mm$^2$ and $L \approx 70$~mm, the fully recruited effective 
modulus $E_f = (k_0 + k_f) L / A \approx 5.6$--11.6~MPa places the fascia lata 
between highly compliant tissues such as skin ($E \sim 0.1$--1~MPa) and stiffer 
structures such as tendons ($E \sim 100$--1500~MPa)~\cite{ristaniemi2021}, 
consistent with its dual role as both a force-transmitting sheath and a gliding 
interface that must accommodate muscle deformation without excessive resistance.

\section{Conclusion}

{This study achieves its primary aim: to develop an experimental and modeling 
framework for the reproducible measurement of the effective viscoelastic properties 
of the fascia lata in vivo, and thereby to provide a quantitative baseline to 
monitor the evolution of fascia lata effective mechanics over time for instance during rehabilitation or in response to training.
The key reside in the combination of a dedicated in vivo ramp--relaxation 
setup with a memory fiber--recruitment--viscoelastic model: neither element alone 
is sufficient --- the experimental protocol preserves the mechanical memory of the 
loading phase within a single continuous test, while the model is the first to 
couple fiber recruitment multiplicatively to dual-timescale viscoelastic elements, 
ensuring that viscous dissipation scales with the load-bearing fiber fraction and 
that the relaxation response retains the full history of the preceding ramp. 
Together, they enable the simultaneous extraction of elastic, viscous, and 
history-dependent parameters from a single in vivo experiment. We emphasize that 
the framework characterizes the effective mechanical response of the fascia lata 
complex including its coupling to the hip--thigh musculoskeletal system, rather 
than intrinsic local material properties. Within this scope, the fitted parameters 
are reproducible within $10\%$ across repeated trials and robust to variations in 
loading amplitude and stretching velocity.}

{Comparison with the literature --- primarily \textit{ex vivo} and \textit{in 
vitro} studies on fascia, tendon, and isolated collagen fibrils, as no in vivo 
reference values for the fascia lata currently exist --- serves two complementary 
purposes. First, it confirms that the fitted viscoelastic parameters, both in 
magnitude and in timescale, are of the expected order for hydrated collagenous 
tissues, lending physical credibility to the measurement framework. Second, and 
building on this agreement, it supports the view that the fascia lata behaves as a 
hierarchical, hydrated composite whose macroscopic response emerges from the coupled 
effects of collagen fiber recruitment, fast collagen--GAG dissipation, and slow 
osmotic fluid redistribution. We stress, however, that these microstructural 
associations remain hypothetical in the absence of local measurements, and identify 
them as the primary target for future experimental investigation combining the 
present protocol with imaging-based or fibril-scale techniques.}

\section*{Conflicts of interest}
There are no conflicts of interest to declare.

%

\end{document}